\documentclass[11pt]{article}
\usepackage{epstopdf}
\usepackage[a4paper,hmarginratio=1:1,vmarginratio=2:3,totalwidth=14.9cm,
 totalheight=22.76cm]{geometry}
 
\usepackage{bm,epstopdf,epsfig,amsmath,amssymb,amsfonts,setspace,
colordvi,latexsym,comment,cancel,verbatim,slashed}

\usepackage{epsfig}
\usepackage{graphicx,graphics}
\usepackage[font=md,captionskip=8pt]{subfig}
\usepackage[usenames,dvipsnames]{color}
\usepackage[noadjust]{cite}
\usepackage{xcolor}  

\usepackage{setspace}
\newcommand{\w}{\omega}
\usepackage[utf8]{inputenc}
\allowdisplaybreaks

\newcommand{\cO}{{\cal O}}

\newcommand{\ra}{\rightarrow}

\newcommand{\be}{\begin{equation}}
\newcommand{\ee}{\end{equation}}
\newcommand{\bea}{\begin{eqnarray}}
\newcommand{\eea}{\end{eqnarray}}
\newcommand{\Ra}{\Rightarrow}

\newcommand{\baa}{\begin{array}}
\newcommand{\eaa}{\end{array}}

\long\def\symbolfootnote[#1]#2{\begingroup
\def\thefootnote{\fnsymbol{footnote}}\footnote[#1]{#2}\endgroup}
\begin{document} 
\begin{flushright}
\end{flushright}
\bigskip\medskip
\thispagestyle{empty}
\vspace{3.8cm}
\begin{center}

{\Large {\bf The fall and the rise of Weyl gauge theory}}

\vspace{1.cm}

 {\bf D. M.  Ghilencea} \symbolfootnote[1]{E-mail: dumitru.ghilencea@cern.ch}

\bigskip
{\small Department of Theoretical Physics, National Institute of Physics
 
and Nuclear Engineering (IFIN) Bucharest, 077125 Romania}
\end{center}

\bigskip
\begin{abstract}\noindent
In 1918 Weyl introduced Weyl conformal geometry  and
its associated quadratic action which was the first gauge theory, of the
Weyl group (of dilatations and Poincar\'e symmetry).
The initial physical interpretation of his theory was however  
short-lived  and  led to the downfall of Weyl geometry as a physical theory.  
We review how this action was re-born into  a  physical Weyl quadratic
gauge theory of gravity.
This is the only (quadratic)  gauge theory of a spacetime symmetry with
a physical gauge boson,  is Weyl anomaly-free, has  {\it exact}  geometric
interpretation,  with all scales of  geometric origin, and generates
Einstein-Hilbert action and a positive cosmological constant  in
its Stueckelberg broken phase. A more fundamental Weyl gauge theory is the
Weyl-Dirac-Born-Infeld  (WDBI) action of Weyl geometry, that is Weyl gauge invariant
in arbitrary $d$ dimensions and that does not need  a UV regulator (!),
of which the (geometrically regularised) Weyl quadratic  gauge theory
is the leading order. For $d\!=\!4$  the WDBI action can
include  SM operators alongside gravitational terms into
a unified description, both geometric and by the gauge principle,
of SM and Einstein-Hilbert gravity, which are recovered in the leading order of
this action.

\end{abstract}

\thispagestyle{empty}

\newpage

\subsection*{[1]. \,\, Weyl conformal geometry and its symmetry}

In 1918 Hermann Weyl introduced   \cite{Weyl1,Weyl1x,Weyl2} the ``true local geometry''
that is now known as Weyl conformal geometry (WG)
which is a generalisation of Riemannian geometry.
He also constructed the action associated
to this geometry, that is quadratic in curvatures,
and presented a physical interpretation of it.
This action was the first gauge
theory, of the  space-time symmetry of dilatations.
While  this geometry was a brilliant construction of
Weyl's genius,  his physical interpretation of this action,
as a unified geometric description of gravity and electromagnetism, was very
short-lived,  due to early criticism from  Einstein \cite{Weyl1} that led to the downfall
of this geometry as a physical theory.

However, Weyl conformal geometry remained to this day 
of great interest to  physicists and mathematicians 
and his work was the foundation of modern gauge theories we have today.
Here we review recent developments that show  that Weyl conformal geometry
with its associated action has a new physical interpretation that actually gives us
more than Weyl  initially thought: a candidate for a
(quantum) gauge theory of gravity, born before quantum mechanics,
that recovers, at large distances, Einstein-Hilbert gravity.

Weyl conformal geometry is defined by classes of equivalence ($g_{\mu\nu}$, $\omega_\mu$)
of the metric $g_{\mu\nu}$  and Weyl gauge field of dilatations $\omega_\mu$,
 related by  a Weyl gauge  transformation
\begin{equation}
  \label{WGS}
  \begin{aligned}
    &\quad  g_{\mu\nu}^\prime=\Sigma^2  \,g_{\mu\nu},\qquad
    \w_\mu'=\w_\mu -    \partial_\mu\ln\Sigma,  \qquad
\Sigma=\Sigma(x)>0.
   \end{aligned}
\end{equation}
%
The definition  is completed by so-called {\it non-metricity} of
WG which has a non-zero $\tilde\nabla_\lambda g_{\mu\nu}$
\begin{equation}
  \label{tildenabla}
\tilde\nabla_\lambda g_{\mu\nu}=- 2  \, \, \w_\lambda\, g_{\mu\nu}, 
  \qquad \textrm{where}\qquad
  \tilde\nabla_\lambda g_{\mu\nu}
  =\partial_\lambda g_{\mu\nu}
  -\tilde\Gamma^\rho_{\lambda \mu} g_{\rho\nu}
  -\tilde \Gamma^\rho_{\lambda \nu} g_{\rho\mu}.
\end{equation}
This differs from metric  (pseudo)Riemannian geometry of Einstein-Hilbert gravity
where  $\nabla_\lambda g_{\mu\nu}\!=\!0$. Assuming
$\tilde\Gamma_{\mu\nu}^\rho=\tilde\Gamma_{\nu\mu}^\rho$ and
using (\ref{tildenabla}) one finds the Weyl connection
$\tilde\Gamma_{\mu\nu}^\rho=\Gamma_{\mu\nu}^\rho+(\delta_\mu^\rho \w_\nu+\delta_\nu^\rho\w_\mu
 -g_{\mu\nu}\,\w^\rho)$ which is invariant under (\ref{WGS}).
 Here $\Gamma_{\mu\nu}^\rho$ is
the Levi-Civita connection;  $\omega_\mu$ measures the deviation 
$\omega_\mu\!\propto \!\tilde\Gamma_{\mu\nu}^\nu\!-\!\Gamma_{\mu\nu}^\nu$;
if $\omega_\mu\!=\!0$ then  $\tilde\Gamma\, (\tilde\nabla_\mu$) becomes
$\Gamma\, (\nabla_\mu)$, respectively,  and WG becomes Riemannian.
 In WG $g_{\mu\nu}$ and $\tilde\Gamma$  are   independent.

 Eq.(\ref{WGS}) defines  {\it Weyl gauge symmetry},
 with $\omega_\mu$  the  gauge boson of this symmetry.
After the later discovery of quantum electrodynamics, it was clear that
Weyl's original physical interpretation of  $\omega_\mu$ as the real photon 
could not work, since electromagnetism corresponds to an
internal $U(1)$ gauge symmetry, not to a space-time (dilatation) symmetry as here.
Hence $\omega_\mu$ is {\it not} the real photon  but a vector field of {\it geometric origin,}
just like the metric, and together with $g_{\alpha\beta}$ make WG  a vector-tensor
theory of gravity, hereafter called Weyl quadratic (gauge theory of)  gravity.

\subsection*{\bf [2].\,\,  ``The fall''}

But a century ago, well before modern gauge theories, Einstein  had
a  different,  powerful argument \cite{Weyl1} against the
physical relevance of Weyl geometry in general or as  a (unified) theory of
gravity and electromagnetism \cite{Weyl1,Weyl1x,Weyl2}, that was  independent
of  whether $\omega_\mu$ was a photon or not. 
The argument was that the {\it non-metricity} of WG 
(i.e. $\tilde\nabla_\mu g_{\alpha\beta}\!\not=\!0$) makes it unsuitable for physics,
because under parallel transport of a vector  (or clock), its length (clock rate)
becomes {\it path-history dependent}.   That means that two identical
atoms  transported along different paths from an initial 
to a final place, end up with different spectral lines, contradicting
experimental evidence (second clock effect)\,\cite{Weyl1}.

This argument lasted a century
and outlived  the attempts to reconsider Weyl geometry and its quadratic gravity
action as a physical  theory. For a
 historical review  of such attempts and  references see~\cite{Scholz}.
Notably, Dirac \cite{Dirac} saw the great potential of this symmetry
in a simplified version of Weyl's theory, linear-only in curvature,
and  constructed a {\it metric} version of it!
Smolin \cite{Smolin} further studied a  similar version, linear in curvature.
But with the advent of modern gauge theories, attention was diverted to 
new symmetries and theories at the time, like Standard Model,
supersymmetry, supergravity or string theory.

\subsection*{\bf [3].\,\,  Modern  interpretation}

Let us set our discussion from the perspective of modern gauge theories.
Much like searches for new physics was based on studying new, internal gauge symmetries
that led to the Standard Model (SM), one can apply the gauge principle  \cite{Noether}
to search for a new 4D space-time symmetry and thus identify the 4D geometry
that may underlie both gravity and SM in a (unified) gauge theory.
The gauged space-time symmetry considered  dictates the actual 4D geometry
and gravity action {\it as a gauge theory action}.
To this purpose, one can gauge: a) Poincar\'e group \cite{Utiyama,Kibble},
b) Weyl  group (Poincar\'e and dilatations)
\cite{Tait,Bregman,CDA2,AC}, c)  full conformal group \cite{Freedman}.
 Case a)  was extensively studied, 
with an unknown UV completion and affected by an infinite series of
higher derivative operators. 
Case c) of gauging the full conformal group leads to conformal gravity \cite{mannheim}.
Strictly speaking this case is {\it not  a true gauge theory} 
since  no dynamical (physical) gauge bosons of dilatation and special conformal
symmetry may be present in the action and spectrum \cite{Kaku}.

This singles out case b) of Weyl gauge theory of (smaller) Weyl group of
Poincar\'e and dilatations, as the only case with a {\it physical}
gauge boson ($\omega_\mu$). By construction this symmetry is realised  
in Weyl geometry \cite{Weyl1,Weyl1x,Weyl2};
this is shown rigourously \cite{AC,Tait} in a tangent space formulation of
the gauged dilatations symmetry, uplifted to space-time by vielbein.
The (quadratic) gauge theory action so obtained in WG is exactly that
written\,by Weyl a century ago \cite{Weyl2}.
And if $\omega_\mu$ is ``pure gauge'' (not dynamical) one recovers the 
particular case of conformal gravity action (plus a conformally coupled scalar)
\cite{review} (section 2.5).

\subsection*{\bf [4].\,\,   Weyl gauge covariance implies metricity}

Let us  see first why the original argument of Einstein
does not apply to Weyl gauge theory. In gauge theories, gauge covariance is
essential to ensure the results are  physical.  In the century-old
formulation ($\tilde\nabla_\mu$) of WG   and its associated action,
the argument was that after  parallel transport,
the length of a vector (or clock rate)  is  changed.  {\it First,}
the  problem with this argument is that this transport (by $\tilde\nabla_\mu$) was
not Weyl gauge covariant, i.e.  it was not physical.
If Weyl gauge covariance is respected, the length of a vector and clock rate 
are indeed invariant (under their parallel transport)
i.e. the theory is metric and there is no path-history dependence 
\cite{Dirac,Ghilen0,Lasenby,non-metricity,CDA,DG1}
Actually, in the symmetric phase of Weyl
gauge theory action there is no mass scale; without a mass scale
there is no clock rate and thus no second clock
effect! {\it Second,} there exists a broken phase,  $\omega_\mu$ becomes massive
(see later) and decouples, so  non-metricity effects are 
strongly suppressed \cite{Ghilen0}.

To detail, notice that  $(\tilde\nabla_\mu + 2 \omega_\mu) g_{\alpha\beta}=0$,
with $q=2$  the Weyl charge of the metric, see (\ref{WGS}).
This suggests that for any tensor $T$ of {\it space-time charge}
$q_T$, in particular for $g_{\mu\nu}$,  that transforms under (\ref{WGS}) like
$T^\prime=\Sigma^{q_T} T$, we can define
a new differential operator $\hat \nabla_\mu$  (replacing
$\tilde\nabla_\mu$)  that, {\it unlike $\tilde\nabla_\mu$}, does transform Weyl-covariantly:
\medskip
\bea\label{qq}
\hat \nabla_\mu T
\equiv (\tilde\nabla_\mu + q_T \, \w_\mu) T\qquad
\Ra\qquad \hat\nabla_\mu' T'=\Sigma^{q_T}\, \hat\nabla_\mu T.
\eea

\medskip\noindent
which  is seen using that $\tilde\Gamma$ is invariant under (\ref{WGS}).
Eq.(\ref{qq}) introduces a Weyl gauge covariant operator
$\hat\nabla_\mu$  by  'covariantisation' of the partial derivative in
$\tilde\nabla_\mu$: $\partial_\mu\ra\partial_\mu +  \rm{charge} \times\w_\mu$.
Since no $\hat\Gamma$ can be associated to $\hat\nabla_\mu$ (because
the charge $q_T$ depends on the field), 
this is {\it not} an affine formulation but it is {\it metric},
since for $g_{\mu\nu}$ we obviously have
\be
\hat\nabla_\mu g_{\alpha\beta}=0.
\ee
Hence Weyl gauge covariance makes Weyl geometry metric.
As a result, if  parallel transport respects Weyl gauge covariance,
the norm of a vector or clock rate remain invariant and the century-old
argument  of Einstein does not apply here.

One can then  define  new  Riemann and Ricci tensors of WG by
using the new operator $\hat\nabla_\mu$, instead of $\tilde\nabla_\mu$ used in the past,
to define the Riemann tensor of WG \cite{CDA}
\be\label{rrr}
    [\hat\nabla_\mu,\hat\nabla_\nu]\, v^\rho=\hat R^{\rho}_{\,\,\,\sigma\mu\nu} v^\sigma
\ee
where  $v^\rho$ is a vector of vanishing Weyl charge in  tangent space.
One then defines the Weyl-Ricci tensor ($\hat R_{\mu\nu}$) and scalar ($\hat R$) curvatures
and we now have a formulation of WG that
is  Weyl gauge covariant {\it in arbitrary $d$ dimensions}.
Indeed, under (\ref{WGS})  \cite{DG1} 
\medskip
\bea\label{tra2}
&&\quad\hat R^\prime=\Sigma^{-2} \hat R,
\qquad\qquad
\hat R^{\prime\mu}_{\,\,\,\,\nu\rho\sigma}=\hat R^\mu_{\,\,\,\,\nu\rho\sigma},
\qquad\qquad
\hat R^{\prime}_{\mu\nu}=\hat R_{\mu\nu},\\
&& \hat\nabla_\mu' \hat R^\prime=\Sigma^{-2}\,\hat \nabla_\mu \hat R,
\qquad
\hat\nabla_\rho' \hat R_{\mu\nu}'=\hat\nabla_\rho \hat R_{\mu\nu},\,\,\,
\qquad\text{etc.}\qquad\qquad\label{wq}
\eea

\medskip\noindent
Thus the curvature tensors/scalar {\it and their derivatives $\hat\nabla_\mu$} 
transform covariantly. This is not  true in the non-metric historical
formulation (of $\tilde\nabla_\mu$), for  derivatives $\tilde\nabla_\mu$
of curvature tensors defined by $\tilde\nabla_\mu(\tilde\Gamma)$.
Also covariant transformations are not possible in Riemannian
geometry where  curvature tensors/scalar transform in a complicated way
under rescaling $g_{\mu\nu}$. This shows the mathematical elegance of WG
in the Weyl covariant   formulation.

We can say that  Weyl conformal geometry
 is a {\it covariantised version} of Riemannian geometry with respect to the
 gauged dilatation symmetry \cite{DG1,review} since 
 $\tilde\Gamma=\Gamma[\partial_\mu g_{\alpha\beta}
 \ra(\partial_\mu+2 \omega_\mu) g_{\alpha\beta}]$.
Various relations between (the squares of) curvature operators
in the Riemannian geometry (of $\nabla_\alpha$), have a similar form in WG in the
Weyl gauge covariant formulation ($\hat \nabla_\alpha$).
Differential and integral calculus 
work similarly, with   $\nabla_\mu\ra \hat\nabla_\mu$ \cite{CDA2}.
The curvatures of WG can be written in terms of their Riemannian version.
For example $\hat R=R-6 \,\nabla_\mu \omega^\mu -6 \omega_\mu\omega^\mu$,
used later on; here $R$ is Ricci scalar  in Riemannian geometry.

\subsection*{\bf [5].\,\, Einstein-Hilbert gravity as broken phase of Weyl gauge theory}

Let us see how Weyl gauge theory recovers Einstein-Hilbert gravity at large distances.
In the Weyl gauge covariant formulation of WG the action is   that
written by Weyl (up to redefinition of couplings)
and is  invariant under (\ref{WGS}) - its simplest version is
\cite{Weyl2,Ghilen0}
\bea\label{Rsq}
S=\int d^4x \sqrt g \,\Big\{
\,\frac{1}{4!\,\xi^2}\,\hat R^2 -\frac{1}{4\,\alpha^2} \hat F_{\mu\nu}^2\,\Big\},
\eea
where $\hat F_{\mu\nu}=\partial_\mu\w_\nu-\partial_\nu\w_\mu$ is the field strength of $\omega_\mu$.
Let us linearise the $\hat R^2$ term in the action,
with  a scalar field $\phi$ by replacing
$\hat R^2\rightarrow -2\phi^2\,\hat R -\phi^4$,
to obtain a classically equivalent Lagrangian
(integrating  $\phi$ via its equation of motion of solution $\phi^2=-\hat R>0$,
recovers $S$).  $\phi^2$ transforms Weyl gauge covariantly just like $\hat R$
(with charge $-2$), so $\ln\phi$ has a shift
symmetry $\ln\phi^2\ra\ln\phi^2-2\Sigma$,
and plays the role of  would-be-Goldstone (``dilaton'' ghost) 
of  gauged scale symmetry (\ref{WGS}).
With this replacement, in the new action one performs a
special scale-dependent gauge transformation (\ref{WGS}) 
with  $\Sigma=\phi/\langle\phi\rangle$, which is fixing $\phi$ to its
vev (assumed to exist); or one simply sets $\phi\ra\langle\phi\rangle$,
and using the relation of
$\hat R$ to its Riemannian counterpart $R$, one finds in {\it Riemannian} notation
\cite{Ghilen0,SMW}
\bea \label{fr}
S=\int d^4x\,\sqrt{g}\,
\,\Big\{-\frac{1}{2}\,M_p^2\, R
-\Lambda\,M_p^2
-\frac14\,\hat  F_{\mu\nu}^2
+\frac12 \, m_\omega^2
\w_\mu\w^\mu
\Big\}.
\eea
where we identify $M_p$ with the  Planck scale, $M_p^2=\langle\phi^2\rangle/(6\xi^2)$,
the cosmological constant $\Lambda=\langle\phi\rangle^2/4$,  and
the mass of Weyl gauge field $m_\omega\!\sim\! \alpha \,M_p$ which is near 
Planck scale (unless one fine-tunes $\alpha$ to ultraweak $\alpha\ll 1$).
Hence, with dimensionful $\langle\phi\rangle$
and dimensionless  couplings $\xi^2\sim\Lambda/M_p^2 \ll\! 1$ and $\alpha\!\sim\! 1$, we 
fixed a similar number of scales in the theory.

We have  an interesting result in (\ref{fr}): the original  Weyl (quadratic) gauge theory of
gravity  (\ref{Rsq}) has a  broken phase, given by
Einstein-Hilbert action and a positive cosmological constant and
the Proca action of 
$\w_\mu$;  this field has become massive by ``absorbing'' the would-be-Goldstone
field $\ln\phi$ (``dilaton'' ghost), via a Stueckelberg mechanism \cite{S}.
The number of degrees of freedom (3) is conserved
as it should in spontaneous breaking (real $\phi$ and
massless $\w_\mu$  were replaced by  massive $\w_\mu$).

The Einstein-Hilbert action  is then a `low-energy'  limit (in unitary gauge)
 of Weyl gauge theory, after massive $\omega_\mu$ decouples 
\cite{Ghilen0}. The  symmetry breaking is accompanied
by a  transition from  Weyl  to Riemannian geometry:
when  $\w_\mu$ decouples, Weyl connection $\tilde\Gamma$ 
is replaced by Levi-Civita connection
(if $\w_\mu\ra 0,$ then $\tilde\Gamma_{\mu\nu}^\lambda\ra \Gamma_{\mu\nu}^\lambda$)
and Weyl geometry becomes Riemannian.
Therefore, the breaking of Weyl gauge symmetry, mass generation
and decoupling of massive $\w_\mu$ have all a {\it geometric interpretation}
as a  transition from Weyl  to Riemannian geometry!
Correspondingly, at large distances one obtains Einstein-Hilbert
action and a cosmological constant $\Lambda\!>\!0$, with
$R=-4\Lambda$ \cite{SMW,Hill}.

All  mass scales were generated by the vev $\langle\phi\rangle$
which has a geometric origin in the $\hat R^2$ term in the action.  Therefore
 these  scales have {\it geometric origin} \cite{non-metricity}. Since $\phi$ is
``eaten'' by $\omega_\mu$ and disappears from the spectrum, there is no
need to stabilize its vev $\langle\phi\rangle$ against quantum corrections as 
required in other theories. This is good news.

This explains  why the mentioned century-old criticism
of  Weyl gauge theory, due to its  ``non-metricity'', is actually avoided:
in the broken phase, the ``non-metricity'' effects due to $\omega_\mu$ are
strongly suppressed by its (large) mass (which is naturally near Planck scale) and
can then be safely ignored - so Weyl's theory is physically viable
\cite{Ghilen0,SMW}.

In Weyl action,  a Weyl-tensor-squared term, $(1/\eta^2)\, \hat C_{\mu\nu\rho\sigma}^2$,
can also be present, with an identical expression as in the Riemannian case
(on  its own it defines conformal gravity \cite{Englert,mannheim}).
In the presence of the Einstein term, it propagates a spin-two ghost state
\cite{Kehagias} with mass again near Planck scale,  $m_\eta\sim\eta M_p$;  for natural
$\eta\sim 1$,  it decouples at low scales and does
not affect predictions. It does affect the unitarity of the theory, but
one does not create such a state in the initial or final state of the
theory \cite{Hawking}. However,  it was recently shown that
this state is {\it not}  present in the spectrum  if
suitable boundary conditions are  imposed for the metric \cite{Maldacena}.
This result is important since then one avoids
unitarity violation (present even in renormalizable
theories of Riemannian gravity \cite{KS}). 

Weyl  action can also have topological terms like
Euler-Gauss-Bonnet term $\hat G$ which is a total derivative in $d=4$
or the parity-odd Pontryagin term \cite{AC}.

\subsection*{\bf [6].\,\, Quantum consistency: no Weyl anomaly in Weyl geometry}

Since Weyl theory  is a gauge theory, the most
important question is whether  Weyl gauge symmetry
is anomaly-free, to ensure that this theory is indeed
consistent at quantum level. Weyl anomaly \cite{A1,A2,A3,A4,A5}
usually plagues
theories with (local) Weyl invariance and has two parts:
one is due to the regularisation  in conflict with classical 
symmetry, another is actually regularisation scale independent.
Let us detail.

First, a regularisation introduces a dimensionful parameter:
a cutoff scale, a subtraction scale $\mu$ in dimensional regularisation (DR), etc.
A DR scheme is preferred since usually it preserves gauge
symmetries by working in $d\!=\!4\!-\!2\epsilon$ dimensions. To
keep couplings dimensionless, a DR scale ($\mu$) is necessary.
But in the special case of Weyl symmetry (local or gauged) this
scale breaks  this symmetry, leading to Weyl anomaly.
One may avoid it  using instead a dynamical scalar
field (dilaton) as regulator {\it field} \cite{Englert,Shaposhnikov};
one then computes (regularised) loop calculations while
preserving this symmetry. When $\phi$ acquires a vev, 
$\mu$ is generated  $\mu\!\!\sim\!\! \langle \phi\rangle$ so\,the
symmetry is broken spontaneously at quantum level
(sometimes $\phi$ is  added by hand to the action, 
changing initial 4D theory).

Second,  Weyl anomaly is more than just a regularisation issue, since it involves
the Euler-Gauss-Bonnet term ($G$). In Riemannian geometry-based theories
with Weyl symmetry, this term is a total derivative in $d=4$, but 
in $d=4-2\epsilon$  this term  gives another anomalous term (Euler term) which is 
regularisation scale  independent. 

How are  these problems avoided in Weyl conformal geometry?
First, a fundamental difference from Riemannian geometry-based theories with
Weyl invariance is that  scalar curvature  $\hat R$ of WG
transforms Weyl gauge covariantly, see (\ref{tra2}).
Thus,  we can have an analytical continuation of  most general Weyl
quadratic gravity  action in $d=4$ into \cite{DG1,review}
\bea\label{W}
S_{\bf w}=\int d^d x
\sqrt{g}\,\,
\Big\{\frac{1}{4!\,\xi^2} \hat R^2 -\frac{1}{4\alpha^2} \hat F_{\mu\nu}^2
-\frac{1}{\eta^2} \,\hat C_{\mu\nu\rho\sigma}^2
+\frac{1}{\rho} \hat G\,\,\Big\}
(\hat R^2)^{(d-4)/4}
\eea
with dimensionless couplings $\xi,\alpha,\eta,\rho$.
With $d=4-2\epsilon$, the factor  $(\hat R^2)^{-\epsilon/2}$  plays the role of DR regulator,
since now all terms in $S_{\bf w}$ have Weyl  gauge  symmetry. This is obvious for
the first three terms, using eqs.(\ref{WGS}), (\ref{tra2}).
Regarding the Euler term, $\hat G$, in WG covariant formulation,
$\hat G=\hat R_{\mu\nu\rho\sigma}\hat R^{\rho\sigma\mu\nu}
-4 \hat R_{\mu\nu}\,\hat R^{\nu\mu} +\hat R^2$ \cite{DG1};
hence it also transforms {\it Weyl gauge covariantly}
in $d$ dimensions, which is a second fundamental difference from Riemannian geometry.
As a result, $\hat G\,(\hat R^2)^{(d-4)/4}\,\sqrt{g}$ in $S_{\bf w}$ is also Weyl gauge invariant.
Hence the action   is now Weyl gauge invariant in $d=4-2\epsilon$ dimensions,
so this symmetry survives at the quantum level and thus there
is no Weyl anomaly \cite{DG1,review}.

The beauty of this regularisation is that
the ``regulator'' $(\hat R^2)^{-\epsilon/2}$  is {\it not} a  scale/field
added by hand to the action, $\hat R$ is part of geometry, in  an
elegant mechanism: {\it geometry itself
does the regularisation} in a Weyl-gauge invariant way!
The breaking of Weyl gauge symmetry proceeds as before;
after massive $\omega_\mu$ decouples (together with
 ``dilaton'' $\ln\phi$),  Weyl connection/geometry become Riemannian  and
Weyl anomaly is restored \cite{DG1,review}.

\subsection*{\bf [7].\,\,   Beyond Weyl gauge theory:  Weyl-Dirac-Born-Infeld action (WDBI)}

There are two natural questions:
first, is there  a more fundamental theory with Weyl gauge symmetry than 
(anomaly-free) Weyl gauge theory of eq.(\ref{W}), that does {\it not} need regularisation?
and  can such fundamental theory  {\it prove} the regularisation used in (\ref{W})?

The answer is yes to both questions -
this is possible in a version of Dirac-Born-Infeld action \cite{BI,Dirac2}
of  Weyl geometry, called Weyl-Dirac-Born-Infeld action (WDBI) \cite{DBI,WDBI}.
The WDBI action goes beyond a {\it quadratic} action and is, in our view,
the most general action in Weyl geometry with Weyl gauge invariance,
valid in arbitrary $d$ dimensions, {\it without
the need of  a UV regulator},  as it is obvious from its expression
 \cite{DBI,WDBI}
\bea\label{WDBI}
S_{\bf WDBI}=\int d^dx \,\sqrt{-\det A_{\mu\nu}^{}},
\qquad\quad
A_{\mu\nu}=
  a_0\,\hat R\,g_{\mu\nu} +a_1 \,\hat R_{\mu\nu} + a_2 \hat F_{\mu\nu},
  \eea
  with {\it dimensionless} coefficients $a_{0,1,2}$, hence
  $A_{\mu\nu}$ has mass dimension 2. From eqs.(\ref{WGS}),(\ref{tra2}),
  $A_{\mu\nu}$ and the action are Weyl gauge invariant in
  $d$ dimensions, in particular in $d=4\!-\!2\epsilon$.

For suitable values of $a_0, a_1,a_2$ \cite{DBI,WDBI} with
$a_{1,2}\ll a_0\!\sim\! 1/\xi^{4/d}$  (recall $\xi^2\sim\Lambda/M_p^2\ll\!1$),
one  expands $S_{\bf WDBI}$ in a power series of $\xi$. Remarkably, 
the leading order ($\xi^0$) of this expansion  {\it is  exactly the regularised} 
Weyl gauge theory action  of eq.(\ref{W}), with identical couplings
\cite{DBI,WDBI}. Hence we actually
     {\it derived}  the {\it geometric} regularisation used in (\ref{W}) and 
the WDBI action is thus more fundamental.
Sub-leading terms of this series,  e.g. $(\hat C_{\mu\nu\rho\sigma}^2)^2/\hat R^2$,
include some quantum corrections  to Weyl action (\ref{W}) \cite{DBI}.
Since  $S_{\bf WDBI}$ is Weyl gauge invariant in 
$d$ dimensions, there is no  Weyl anomaly.

Briefly, the WDBI action is an elegant generalisation of
Weyl (quadratic) gauge theory of gravity and
 unique among gauge theories since it does {\it not} need a
 regularisation, while it remains Weyl gauge invariant
 for arbitrary $d$ (for $d=4$  one
 only needs pure  analytical continuation by replacing
 $d=4\ra d=4-2\epsilon$).
 To appreciate this, note that 
 not even in string theory can Weyl symmetry be respected
 by regularisation/at quantum level
(on the  2D Riemannian worldsheet, not in physical space-time as here) {\it without
  imposing an additional} (sufficient) condition, of  vanishing
beta function,  leading to Einstein equation and the need for $d=10$ dimensions.
Comparatively, $S_{\bf WDBI}$ here  is Weyl gauge invariant for any  $d$,
but $d=4$ is good enough for all our purposes.

\subsection*{\bf [8].\,\,   Weyl gauge symmetry and SM}

What about adding matter to Weyl geometry?
SM  with a vanishing Higgs mass {\it parameter} ($m_H\!=\!0$) is scale invariant.
This may suggest that this symmetry is more fundamental, thus one can gauge it.
This is realised by embedding SM  in WG, hereafter called SMW;
the embedding is natural, without new degrees of freedom beyond  WG and SM
\cite{SMW}.

At classical level, the action of SM gauge bosons and fermions (including Yukawa sector)
 is already (locally) Weyl invariant in (pseudo)Riemannian
 geometry. Hence, this action remains identical when SM is embedded
  in (d=4) Weyl geometry
\cite{SMW}.
Regarding  the Higgs, it ``knows'' about mass, so the   Higgs action is mildly
changed when embedded in WG, and $H$ couples to $\omega_\mu$.
There is a  term $H^\dagger H \hat R\sqrt{g}$ ($\hat R$ depends
 on $\omega_\mu$), while the Higgs kinetic term
 is upgraded to be Weyl gauge invariant, $\vert\hat \nabla_\mu H\vert^2\sqrt{g}$, with
$\hat\nabla_\mu H=(D_\mu -\alpha \,\omega_\mu)H$, and $D_\mu H$ the
 SM covariant derivative. So the Higgs Lagrangian is upgraded to
 $\sqrt{g}\,\, \big[ \vert\hat \nabla_\mu H\vert^2-(1/6)\, \xi_h \,H^\dagger H \hat R 
   -\lambda\vert H\vert^4\big]$. This ends our brief review of the SMW action.

The breaking of Weyl gauge symmetry in SMW proceeds  as before,
 with minor corrections from the SM Higgs sector.
 Studies show that the coupling of $H$ to $\omega_\mu$ is not
 relevant experimentally \cite{SMW}.
But there is an indirect  prediction of SMW that
can be tested experimentally and  is important: due  to the $\hat R^2$
term in action, SMW has a  successful, (gauged version of)
Starobinski-Higgs inflation  \cite{WI3,WI1}.
Good fits for  galactic rotations curves are also possible,
with a Weyl geometric ($\omega_\mu$)  solution for  dark matter \cite{Harko}.

\bigskip\noindent
{\bf $\bullet$ WDBI action of SM}

\bigskip\noindent
Remarkably, one can extend the
WDBI action (\ref{WDBI})  to have  SM operators alongside
gravitational terms  in  $A_{\mu\nu}$  of action  (\ref{WDBI}) \cite{WDBI}.
One  thus extends  $A_{\mu\nu}$ of (\ref{WDBI}) by including dimension-two operators
built from both SM and WG operators, on equal footing, and that are SM and
Weyl gauge invariant. The new action is (below $d=4-2\epsilon$)
\be
S^\prime_{\bf WDBI} =\int d^d x \sqrt{-\det \,A_{\mu\nu}^\prime},
\ee
with \cite{WDBI}
\begin{align}
  A^\prime_{\mu\nu}
  &= a_0\, \hat R g_{\mu\nu}+ a_1\,\hat R_{\mu\nu} + a_2 \hat F_{\mu\nu}
  + a_3 \, F_{\mu\nu}^{(1)}
\nonumber\\
  &+  g_{\mu\nu} \,\big[\,
    a_4^{(j)} \,F^{(j)\,2}_{\alpha\beta} \hat R^{-1}
+ a_5\,\vert \hat\nabla_\alpha H\vert^2\, \hat R^{1-d/2} 
+a_6 \,\vert H\vert^2 \hat R^{2-d/2} 
+a_7 \vert H\vert^4\,\hat R^{3-d}
 \nonumber\\
&+a_8\, \big(
  i\, \overline\psi \gamma^a\,e_a^\alpha\hat\nabla_\alpha\psi+\textrm{h.c.}
  \,\big) \, \hat R^{1-d/2}\,
+a_9\,\big( 
\overline \psi_L\,Y_\psi\, H\psi_R
+\overline \psi_L Y^\prime_\psi\tilde H\,\psi^\prime_R+\textrm{h.c.}\big)
\, \hat R^{2-3\,d/4}\,
\nonumber\\
&+ a_{10}\, \hat F_{\alpha\beta} \hat F^{\alpha\beta} \hat R^{-1} 
+a_{11}\, \hat F_{\alpha\beta}  F^{(1)\,\alpha\beta} \hat R^{-1}
\big]. 
    \label{amunu}
\end{align}

\medskip\noindent
$a_{0,1..11}$ are dimensionless coefficients,
$F_{\mu\nu}^{(j)}$: SM field strengths $j\!=\!1,2,3$,  ($F_{\mu\nu}^{(1)}$ for
$U(1)_Y$).

Broadly speaking, the  metric $g_{\mu\nu}$, gauge bosons $F_{\mu\nu}^{(1)}$, SM matter, 
all contribute on equal footing to $A_{\mu\nu}^\prime$,  then  $A_{\mu\nu}^\prime$
can be regarded as the ``new'' space-time
metric and $d^dx \sqrt{\det(A_{\mu\nu})}$  is the new space-time volume.
Since  $d=4$ this new WDBI action that includes SM,
is background independent; gravity is not a background in which SM
exists, this action
with $d=4$ makes no  distinction between the ``brane'' (matter) and the
``bulk'' (gravity) that exists in string theory (where $d=10$).
In the  sense
of contributions to the new ``metric'' ($A_{\mu\nu}^\prime$), our action gives  a
geometric unification of SM and gravity,
removing a distinction matter-geometry.
Such unification also exists in the
gauge sense: external (gauged dilatations) and internal (SM)
gauge symmetries act on equal footing to generate the action.
This action has a vague similarity to a D3-brane effective
action but that has no Weyl gauge symmetry that
was so important here to generate mass, as we saw.

This new $A'_{\mu\nu}$ generates the  WDBI gauge theory action 
of  WG {\it ``plus''} SM interactions, that is  general, beyond
familiar quadratic actions in gauge theories.
The elegance  of this  WDBI action of WG+SM is seen
if we expand  again  $[-\det A'_{\mu\nu}]^{1/2}$  in powers of  $1/a_0\!\sim\! \xi\ll\!1$:
we find in leading order ($\xi^0$)  the {\it regularised}  Weyl quadratic
gravity action of (\ref{W})  {\it plus}  the
Weyl gauge invariant {\it regularised} action of SM 
\cite{WDBI} (below $d\!=\! 4-2\epsilon$)
\smallskip
\begin{align}\label{f}
\!\!\!S_{\bf d}=\int d^dx\,&\sqrt{g}\,
\,\Big\{
  (\hat R^2)^{(d-4)/4}
\Big[\,\frac{1}{4!\,
    \xi^2} \,\hat R^2
 -\frac{1}{4\alpha^2}\,\hat F_{\mu\nu}^2
  -\frac{1}{\eta^2}   \hat C_{\mu\nu\rho\sigma}^2+\frac{1}{\eta^2}\hat G
     -\frac{1}{4\alpha_j^2} \,F^{(j)}_{\mu\nu} F^{(j)\,\mu\nu}
    \Big]
 \nonumber\\[-3pt]
  &\, + \vert\hat\nabla_\mu H\vert^2 -(1/6)\, \xi_H\vert H\vert^2\,\hat R
 -\lambda\,\vert H\vert^4\,\hat R^{2-d/2}
+ \big[
  (i/2) \,\overline \psi_L \gamma^a e_a^\alpha \nabla_\alpha\psi_R
  +\textrm{h.c.}\big]
\nonumber\\
 &\, + \big(\overline\psi_L Y_\psi H\psi_R
+ \overline\psi_L\, Y_\psi^\prime \tilde H\,\psi^\prime_R
+   \textrm{h.c.}\big) \,\,\hat R^{1-d/4}
\!+\cO\big[1/\hat R^3\big]\Big\}
+\cO(\xi),
\end{align}

\smallskip\noindent
for suitable  values of coefficients $a_{0,1,...11}$
as functions  of physical couplings  in eq.(\ref{f}) \cite{WDBI}.

Weyl geometry and the action $(-\det A'_{\mu\nu})^{1/2}$
generated  automatically a unique regularisation not only for the $d=4$ Weyl
action (as  in (\ref{W})), but now also for the $d\!=\!4$
SM action, while respecting both SM and
Weyl gauge symmetry in $d\!=\!4\!-\!2\epsilon$. 
Given this symmetry, the theory is  Weyl-anomaly free.
We see again that for  a $d=4$ WDBI action of WG+SM all one has to do is 
to analytically continue   $d=4 \ra d\!=\!4\!-\!2\epsilon$,
but {\it no UV regulator}  is needed.
The  Einstein-Hilbert action and  $\Lambda\!>\!0$
are recovered as before (in the absence of SM), together with SM action,
in the spontaneous broken phase of Weyl gauge symmetry.
Riemannian geometry is restored below $\sim M_p$, after massive $\omega_\mu$ decouples.
The leading order of (\ref{f}) is the  SMW action of SM and quadratic gravity in WG.

\subsection*{[9].\,\,\,  Conclusions}

The original Weyl  gauge theory
 (of Poincar\'e and dilatations symmetry)
is re-born as  a quadratic gauge theory of gravity that
is Weyl-anomaly-free,  with  {\it exact} geometric interpretation;
 all mass scales (Planck, $\Lambda$, $m_\omega$)
have similar  geometric field origin, generated by Stueckelberg
 mechanism. In the broken phase  Einstein-Hilbert gravity is recovered
together with  $\Lambda\!>\!0$.
A more general Weyl gauge theory of gravity exists,
called Weyl-Dirac-Born-Infeld gauge theory (WDBI) of Weyl geometry.
The WDBI action is unique - it
does {\it not} need  a UV regulator (!), it is
mathematically well-defined and Weyl gauge invariant in {\it arbitrary} $d$ dimensions,
hence there is no Weyl-anomaly.
When  expanded in  power series in $d\!=\!4-\! 2\epsilon$ the geometrically
regularised  Weyl quadratic gauge theory of gravity 
is the leading order!
This WDBI  action can  include  SM contributions (if $d\!=\!4$)
in a unified description, both geometric and
by the gauge principle, of Einstein-Hilbert gravity and SM.
The WDBI action is thus a great candidate for\,a\,quantum gravity theory.

{
}
\end{document}